\documentclass[twocolumn,showpacs]{revtex4}    
\usepackage{graphicx}%
\newcommand{\figwidth}{3.0in} 
\begin{document}
\title{Commensurate-incommensurate transitions in quantum films: \\
Submonolayer molecular hydrogen on graphite}
\author{Kwangsik Nho} 
\affiliation{Center for Simulational Physics, University of Georgia,
 Athens, Georgia, 30602-2451, U.S.A}
\author{Efstratios   Manousakis}
\affiliation{Department   of  Physics   and  MARTECH,  
Florida  State University,
Tallahassee, Florida 32306, U.S.A and \\
Department of Physics,  University of Athens,
Panepistimiopolis, Zografos, 157 84 Athens, Greece
}
\date{\today}
\begin{abstract}
We have used the Path Integral Monte Carlo method to simulate a monolayer
of molecular hydrogen on graphite above 1/3 submonolayer coverage.
We find that at low temperature and as the coverage increases
the system undergoes  a series of transformations starting 
from the $\sqrt{3} \times \sqrt{3}$ commensurate  solid near 1/3 
coverage. First, a phase is formed which is characterized by a 
uniaxially compressed incommensurate solid with additional mass 
density modulations along the same direction which can be viewed as 
an ordered domain-wall solid with a characteristic domain-wall distance 
which depends on the surface coverage.
At low temperature and higher coverage there is a transition to 
an incommensurate solid which is rotated relative to the 
substrate commensurate lattice.
As a function of temperature the domain-wall ordering first melts into 
a fluid of domain-walls  and at higher temperature the solid 
melts into a uniform fluid. Regardless of the large amplitude of quantum 
fluctuations, these phase transitions are analogous to those in classical 
monolayer films.  Our calculated values for the surface coverage and 
temperature where these transitions occur, the calculated structure factors
and  specific heat are in general agreement with the available
experimental results with no adjustable parameters.

\end{abstract}

\maketitle

\section{Introduction}

The commensurate-incommensurate transition has been expensively studied
in classical monolayer films both 
theoretically and experimentally. Monolayers of hydrogen or
helium are quantum mechanical systems and, in principle, one might suspect
a different behavior due to the effect of strong quantum fluctuations.

\begin{figure}[htp]
\includegraphics[width=\figwidth]{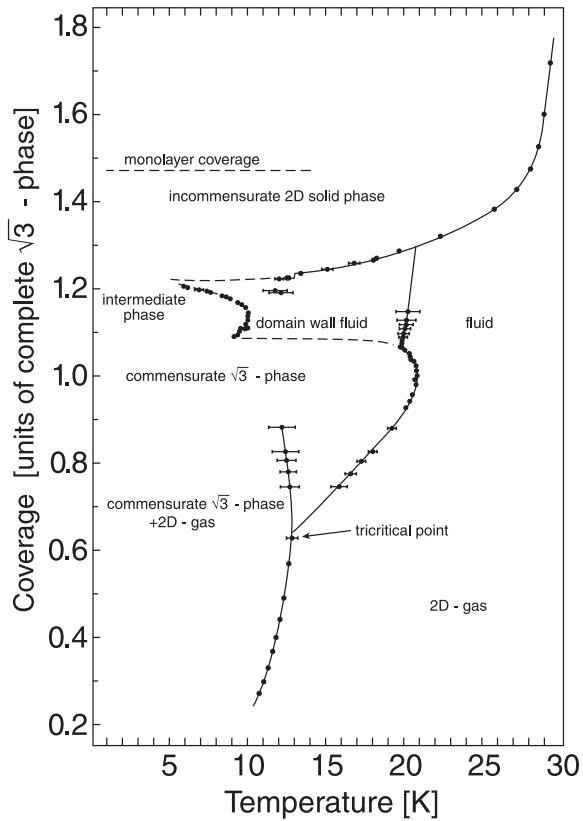}
\caption{Phase diagram of  molecular hydrogen
adsorbed on graphite from Ref.~\cite{Wiechert91a} reproduced
here for easy reference.
Density 1.0 corresponds to the complete commensurate $\sqrt{3}\times\sqrt{3}$
  phase.}
\label{fig-phase}
\end{figure}

The monolayer phase diagram of molecular para-hydrogen adsorbed on 
graphite as inferred from experimental studies\cite{Freimuth87} is  
shown in Fig.~\ref{fig-phase}.  
This phase diagram was originally drawn based on the anomalies found in 
the specific heat\cite{Motteler,Chaves,Freimuth85}; for the characterization
of the various phases, low energy electron 
diffraction (LEED)\cite{Cui86,Cui}
as well as neutron diffraction\cite{Nielsen77,Freimuth87,Wiechert91a}
studies have been carried out. At 1/3 coverage the molecules condense
on the surface of graphite in 
a $\sqrt{3}\times\sqrt{3}$ commensurate solid.
In the low coverage region
$(\rho < 0.6)$ (in units of the $\sqrt{3}\times\sqrt{3}$ 
commensurate density), it forms a commensurate solid-gas 
coexistence phase at low temperature
and a 2D gas phase at higher temperature ($T > 10 K$). For coverages 
$0.6 ^{<}_{\sim} \rho ^{<}_{\sim} 0.9$,
as a function of temperature, there is a transition 
from a commensurate solid cluster phase to 
a commensurate solid phase with vacancies at higher temperature and 
to a 2D gas at even higher temperature. At density somewhat higher than 
unity the so-called $\alpha$-phase is formed which is believed to
be a striped domain-wall solid phase  and at higher temperature
it transforms to the so-called $\beta$-phase  
and to a fluid phase at even higher temperature.
At high densities a transition take place to a triangular incommensurate 
solid phase. 

The phase diagram of molecules formed from  the isotopes of the
hydrogen such $D_2$ and $HD$ molecules physiorbed on graphite has
also been studied\cite{Freimuth90,Lauter90} and it 
is similar to that of $H_2$ on graphite but more complex.

We have recently studied\cite{NM} the phase diagram of molecular hydrogen 
on graphite at and below 1/3 coverage using 
Path Integral Monte Carlo (PIMC) simulation.
Our computed phase diagram was in general agreement with 
that inferred from the experimental studies for this coverage
range. In this paper we extend the PIMC calculation to study 
the phase diagram above the commensurate density which is less 
well understood both theoretically and experimentally.
At coverages above $\rho \simeq 1.05$
the monolayer undergoes a commensurate-incommensurate(C-IC) transition. 
While the existence of the so-called $\alpha$ and $\beta$ phases
were known from the specific heat anomalies, 
their characterization came from  LEED\cite{Cui86,Cui} and 
neutron scattering\cite{Freimuth87} studies.
Using LEED
experiments Cui and Fain\cite{Cui86,Cui} observed  
a uniaxial IC solid with striped superheavy domain walls
 and a rotated triangular IC solid at higher coverages
$\rho \geq 1.23$.  Freimuth, Wiechert, and Lauter\cite{Freimuth87} (FWL)
presented a neutron diffraction study of the C-IC transition  and
their results are in agreement with the LEED results.
They examined the  commensurate-incommensurate solid 
transition, especially the striped domain-wall solid phase.
In the $\alpha$-phase the diffraction intensity has a main peak 
characteristic of a  
compressed lattice and a satellite peak
on the lower side of the 
commensurate peak wave-vector position $k_{c}=1.703\AA^{-1}$
that arise from the spatial modulation
due to ordered striped-domain walls.
As coverage increases, the separation 
of the two peaks increases and the satellite intensity decreases.
At higher temperatures the peak height drops 
and the satellite peak vanishes which implies that the commensurate 
domains vanish and the system becomes an isotropic fluid phase.
As coverage increases, the first molecular hydrogen layer forms a rotated  
incommensurate solid (RIC) phase. 
Neutron diffraction studies show a sharp and intense peak at wave-vector 
$k=1.97 \AA^{-1}$ at the density $\rho=1.34$
which reveals an IC equilateral triangular structure. This phase is 
continuously compressed as the 
coverage increases up to the highest density allowed before layer promotion.

In parallel to these experimental investigations several important
theoretical studies were undertaken to understand the 
commensurate-incommensurate
 transition\cite{novaco-mctague,vilain,bak,Copper,Timothy}
in physiorbed systems. The phases and the phase transitions which were 
theoretically predicted and studied can be understood as classical
phenomena due to the strain induced on the adsorbate molecules of the
monolayer film due to the substrate periodic structure
which arise from the competition between adsorbate-adsorbate and 
adsorbate-substrate interactions. Molecular hydrogen and atomic helium 
physisorbed on graphite are quantum films characterized by 
strong zero point motion. Therefore, one could question the degree
to which a classical picture might be valid and might expect new
phenomena and phases to occur.  Motivated by these thoughts
we extended our earlier investigation\cite{NM} to study this system
above 1/3 coverage up to layer completion 
starting from the known hydrogen
molecule-molecule and hygrogen-graphite 
interactions\cite{Silvera,Crowell,Steele,Carlos} and
using
the  PIMC\cite{Ceperley95} which is a 
Quantum Monte Carlo technique adopted for the study of strongly interacting
quantum films by Pierce and Manousakis\cite{Pierce}.

We have simulated the first layer of molecular hydrogen on graphite with 
a variety of simulation cells
that are appropriate for examining different phase regions of the phase 
diagram.
We have applied periodic boundary conditions along both directions 
parallel to the surface of the substrate.
We have computed expectation values for the total energy, the static structure factor, the probability distribution, and the specific heat by means of 
the path-integral Monte Carlo simulation method
using the multi-level metropolis method. In the multi-level bisection method
we have used level 3 which optimizes the  acceptance ratio. 
To thermalize the system, we typically carried out of the order of 
15,000 MC steps and we carried out of the order 
of 2,000-20,000 MC steps in order to compute observables. 

\begin{figure}[htp]
\includegraphics[width=2.0 in]{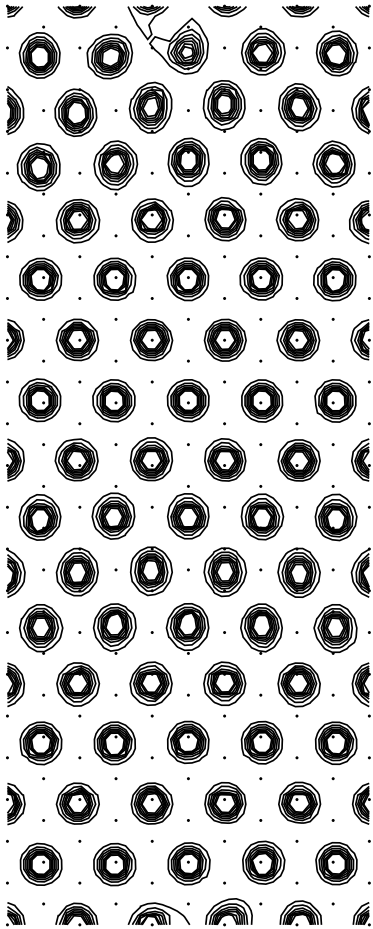}
\caption{Contour plot of the probability distribution at the density
0.0694 $\AA^{-2}$ and $T$= 1.33 $K$. 
The simulation cell is 21.295 $\AA$ $\times$ 54.098 $\AA$ (80 
$H_{2}$ molecules).
The filled circles indicate graphite adsorption sites.
The solid structure is uniaxially compressed along our y direction.
In addition, two commensurate solid domains separated by the 
incommensurate solid domains can be seen which implies a mass density
wave along the same direction.}
\label{fig-80fw}
\end{figure}

\begin{figure}[htp]
\includegraphics[width=2.0 in]{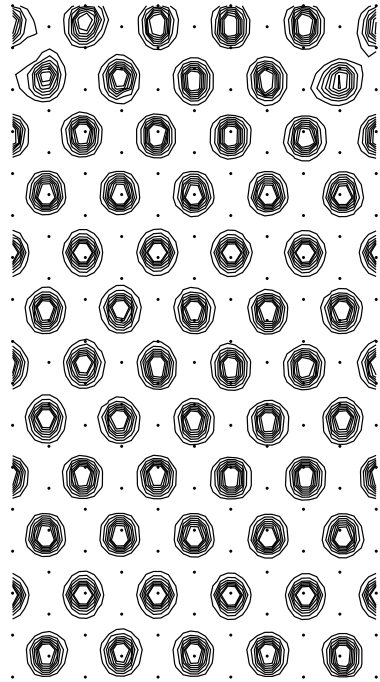}
\caption{Contour plot of the probability distribution at the density
$0.0716 \AA^{-2}$ for $T$= 1.33 $K$. 
The simulation cell is 21.295 $\AA$ $\times$ 39.344 $\AA$ (60 molecules).
The filled circles indicate graphite adsorption sites.
The solid structure is uniaxially compressed along our y direction.
The striped domain-wall solid phase can be seen.}
\label{fig-60fw}
\end{figure}

\section{Uniaxially Compressed Solid With Ordered Domain Walls}

In this section we discuss the region above the commensurate coverage
where experimentally the ordered domain-wall solid was found.
We carried out a simulation at two densities 
in this region, the same densities discussed by FWL, where they inferred
the existence of relaxed superheavy domain walls.

We used simulation cells 
that can accommodate 
the structures that have been proposed by FWL.
a) First, we consider $\rho= 0.0694 \AA^{-2}$. We have chosen 
a simulation cell with dimensions $x$= 5$\sqrt{3}$ $a_{gr}$ and 
$y$= 22 $a_{gr}$, where $a_{gr}=2.459 \AA$ is the carbon-carbon 
distance on the graphite surface and  80 hydrogen molecules 
to produce the above density. 
In Fig.~\ref{fig-80fw} we give the contour plot of the calculated
probability distribution where we see that two 
commensurate-solid domains are 
separated by the incommensurate solid domains. The solid structure is
uniaxially compressed along our $y$ direction such that a new row of molecules
is introduced for every 8 rows. 
Notice that  the period in the $x$-direction is 
$\sqrt{3}a_{gr}$  and  there is modulation along the $y$ direction
with period $ 11 a_{gr} $, so that the wavelength $\lambda_{s}$ of
this striped domain-wall modulation is $\lambda_{s} = 27.049$.
b) Second, we have performed the simulation at the density 
$\rho = 0.0716 \AA^{-2}$.
We have chosen a simulation cell with dimensions $x= 5 \sqrt{3} a_{gr}$
and $y= 16 a_{gr}$ and 60 hydrogen molecules in order to 
achieve the above mentioned density. 
The calculated contour plot of the probability distribution 
is shown in Fig.~\ref{fig-60fw}. 
Notice that in this case also there is an ordered striped domain-wall 
solid phase along our $y$ 
direction. Namely, the amplitude of the molecular density wave
is modulated along the $y$ axis with a period of about half
our  cell size along the $y$ axis. 
Notice that there are two commensurate 
solid domains separated by denser regions.

\begin{figure}[htp]
\includegraphics[width=\figwidth]{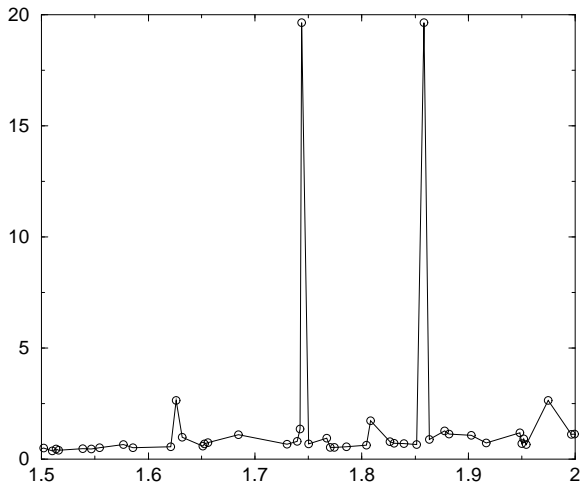}
\caption{Static structure factor $S(k)$ at the density
 $0.0694 \AA^{-2}$ (and corresponds to Fig.~\ref{fig-80fw}) for $T$= 1.33 $K$
as a function of the magnitude of $\vec k$. 
The main Bragg peaks are at $k_{1}$= 1.743 $\AA^{-1}$ and 
$k_{2}$= 1.858 $\AA^{-1}$. There is a satellite peak 
located at  $k_{sat}^{c}$= 1.626 $\AA^{-1}$.}
\label{fig-s80fw}
\end{figure}

\begin{figure}[htp]
\includegraphics[width=\figwidth]{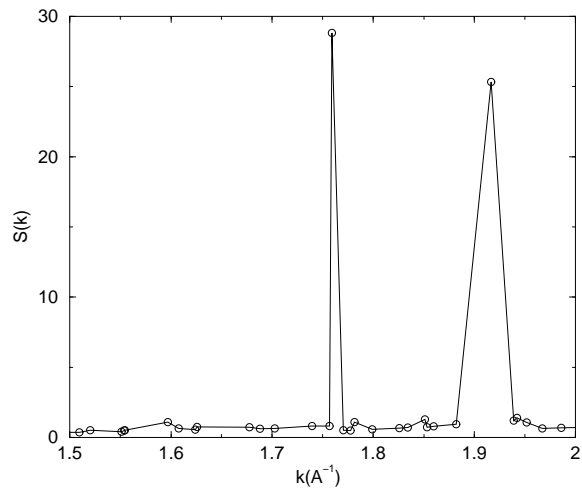}
\caption{Static structure factor $S(k)$ at the density
 $0.0716 \AA^{-2}$ (and corresponds to Fig.~\ref{fig-60fw}) for $T$= 1.33 $K$. 
The main Bragg peaks are at $k_{1}$= 1.759 $\AA^{-1}$ and 
$k_{2}$= 1.916 $\AA^{-1}$. Notice that the satellite peak 
at $k_{sat}^{c}$= 1.597 $\AA^{-1}$ is significantly weaker than the
satellite peak at the lower density shown in Fig.~\ref{fig-s80fw}.}
\label{fig-s60fw}
\end{figure}

Our computed static structure factor $S(k)$ for 
$\rho=0.0694  \AA^{-2}$ is shown in Fig. \ref{fig-s80fw}. 
The main Bragg peaks of this structure occur at
${\bf k}_{1}$ = (1.475 $\AA^{-1}$,0.929 $\AA^{-1}$) and 
${\bf k}_{2}$ = (0,1.858 $\AA^{-1}$). The satellite peak occurs 
at ${\bf k}^{c}_{sat}$= (0,1.626 $\AA^{-1}$). 
The experimental values of the magnitude of the wave-vectors at the peaks
are  $k_{main}^{exp}$= 1.743 $\AA^{-1}$ and $k_{sat}^{exp}$= 1.632 $\AA^{-1}$,
which compare well with our computed values of $k_{1}^c$= 1.743 $\AA^{-1}$ and 
$k_{sat}^c$= 1.626 $\AA^{-1}$.
We believe that FWL could not observe the peak at ${\bf k}_{2}$
because of the strong interference with the (002) graphite peak.

The interpretation of these results is as follows:
Analyzing the contour plot of Fig.~\ref{fig-80fw} we find that 
a) the solid appears uniaxially compressed along the $y$ direction, 
by adding another row  for every 10 rows of molecules and spreading them 
evenly, while
along the $x$ direction the average spacing between the molecules
remains that of the commensurate solid.
b) superimposed there is a density modulation along the $y$ direction
which has wavelength several times greater than the average nearest neighbor 
distance but of small amplitude. The actual size of the wavelength $\lambda_s$
of this striped domain-wall modulation 
is half of the length of our simulation cell in the $y$-direction,
i.e., $\lambda_s = 27.05 \AA$, as discussed in the previous paragraph.
The  two main  Bragg peaks  correspond to the unit vectors which span  
the reciprocal space of this  uniaxially compressed
triangular solid  structure. 
The satellite peaks are due to lateral modulation in our $y$ direction and
are  located  at  ${\bf k}_{sat}$  = ${\bf k}_{1,2}-{\bf k}_s$.
${\bf k}_{1,2}$ correspond to the wave vector  of each of the two main Bragg
peaks and ${\bf k}_s  =  (0,2 \pi /\lambda_s)$, where $\lambda_s$ is the
wave length of a unit cell for the striped domain-wall solid structure
in the modulated direction.
Using this $\lambda_{s}$ value, ${\bf  k}_2 = (0,1.858 \AA^{-1})$ 
and ${\bf k}_s =  (0,0.232 \AA^{-1})$ (obtained by using the values
of $\lambda_s=27.049 \AA$ found by inspection of Fig.~\ref{fig-80fw})
we can find that the satellite peak position is
at ${\bf  k}_{sat} = (0,1.626\AA^{-1})$, which agrees well with our 
peak value.

The structure factor for the case of $\rho= 0.0716 \AA^{-2}$ 
(Fig.~\ref{fig-60fw}) is shown in Fig.~\ref{fig-s60fw}.
The main peaks are
${\bf k}_{1}$ = (1.475 $\AA^{-1}$,0.958 $\AA^{-1}$) and  
${\bf k}_{2}$ = (0,1.916 $\AA^{-1}$) which correspond to 
the uniaxially compressed solid produced by adding another row of
molecules for every 6 rows of molecules and spreading them 
evenly. This, in addition, 
forms a superscructure with wavelength $\lambda_s = 19.672 \AA$ 
which gives a satellite peak at ${\bf k}^{c}_{sat}$= (0,1.597 $\AA^{-1}$).
Notice that the scattering intensity, relative to the previously
discussed case of $\rho =0.0694 \AA^{-2}$, has the peaks shifted at
higher values of $k$ as expected and the intensity of the
satellite peaks is decreased as in the experiments.

\section{Domain Wall Melting}

\begin{figure}[htp]
\includegraphics[width=\figwidth]{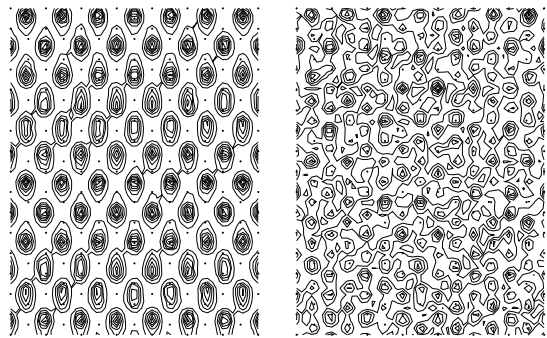}
\caption{Contour plot of the probability distribution 
at the density $0.0716 \AA^{-2}$ at two temperatures 11.11 $K$ 
and 16.67 $K$.}
\label{fig-pd0716}
\end{figure}

In Fig. \ref{fig-pd0716} we show the contour plot of the the probability 
distribution where a striped domain-wall fluid phase (at $T=11.11 K$) and
a fluid phase (at $T=16.67$) are evident. Notice that the stripe domain-wall
fluid phase (left part of Fig.~\ref{fig-pd0716}) is characterized by
mobile commensurate and incommensurate domains. To understand
this contour plot we need to be reminded of the
periodic boundary conditions used in our simulation and the fact that
these domains cannot intersect each other. This implies that one
domain will oscillate back and forth between the neighboring domains.
We can determine the melting temperature of the domain-wall solid phase
from the temperature dependence of the specific heat and from the temperature
dependence of the peak height of the static structure factor $S(k)$. 

\begin{figure}[htp]
\includegraphics[width=\figwidth]{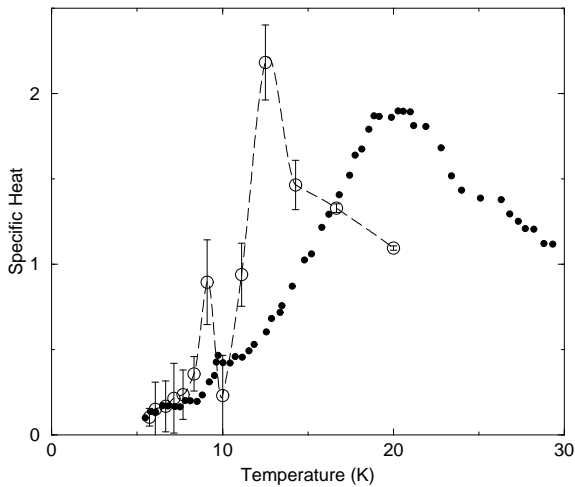}
\caption{Specific heat versus temperature at the density
$0.0716 \AA^{-2}$. The long-dashed line is a spline fit to the specific
heat values and is a guide to the eye. The filled circles are the experimental specific heat at the density $0.0716 \AA^{-2}$.
The simulation cell is 29.813 $\AA$ $\times$ 39.344 $\AA$ (84 $H_{2}$ 
molecules).
Our computed values for the melting and evaporation temperature
 are the peak positions, 
9.09 $K$ and 12.5 $K$, respectively.}
\label{fig-spec}
\end{figure}

\begin{figure}[htp]
\includegraphics[width=\figwidth]{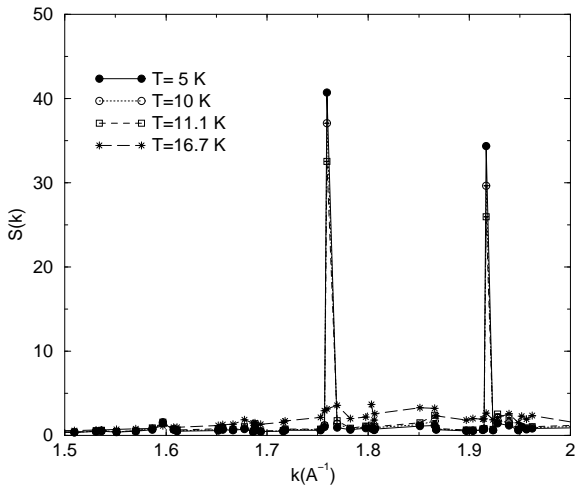}
\caption{Static structure factor $S(k)$ at the density $0.0716 \AA^{-2}$
for various temperatures.}
\label{fig-sk0716}
\end{figure}

\begin{figure}[htp]
\includegraphics[width=\figwidth]{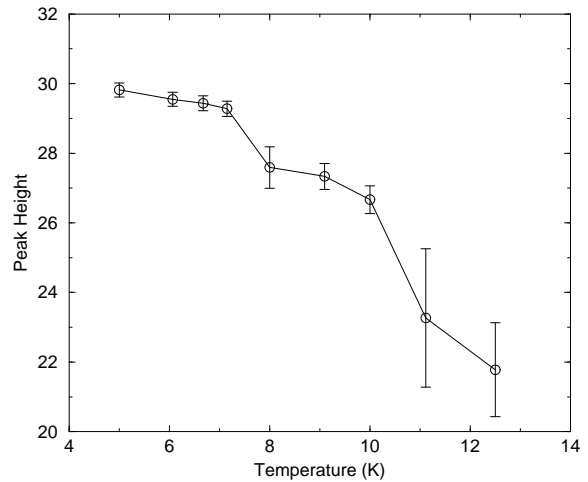}
\caption{The peak height of the static structure factor $S(k)$ 
versus temperature
at the density $0.0716 \AA^{-2}$.
The simulation cell is 21.295 $\AA$ $\times$ 39.344 $\AA$ 
(60 molecules).}
\label{fig-peakheight}
\end{figure}

The calculated specific 
heat as a function of temperature is shown in Fig.~\ref{fig-spec} and
  is characterized by two peaks, 
the first corresponds to the melting of the domain-wall solid while 
the second peak indicates the melting of the stripes and the solid into a
fluid.
Our computed  values for the melting temperature are generally lower
that the experimental values.
For example, for $\rho=  0.0716 \AA^{-2}$  we find that at $T = 9.09 K$ 
the striped domain-wall solid phase undergoes 
a transition to the domain-wall fluid phase and at $T = 12.5 K$ the 
$\beta$-phase becomes an isotropic fluid phase. 
Factors for obtaining lower values for the critical
temperature than the experimental values could be the finite-size effects and
the interaction strength used to describe the hydrogen-graphite interaction. 
We also computed the static structure factor shown in Fig.~\ref{fig-sk0716} 
and 
we studied the temperature dependence of the height of its 
first main peak which is shown in Fig.~\ref{fig-peakheight}.
The peak height significantly decreases near the melting temperature 
of the domain-wall solid  and near the domain-wall evaporation temperature.

\section{Rotated Incommensurate Solid}

\begin{figure}[htp]
\includegraphics[width=\figwidth]{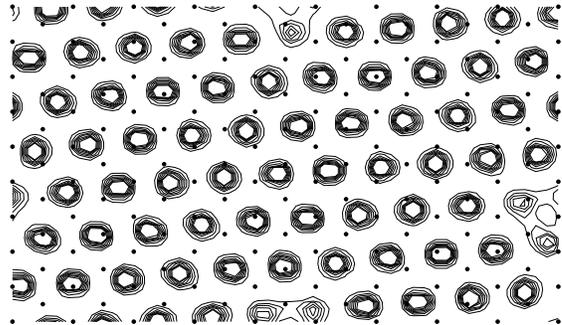}
\caption{Contour plot of the probability distribution at the density 
$0.0849 \AA^{-2}$. The simulation cell is 38.331 $\AA$ $\times$ 
22.131 $\AA$ (72 $H_{2}$ molecules).}
\label{fig-ic2fw}
\end{figure}

\begin{figure}[htp]
\includegraphics[width=\figwidth]{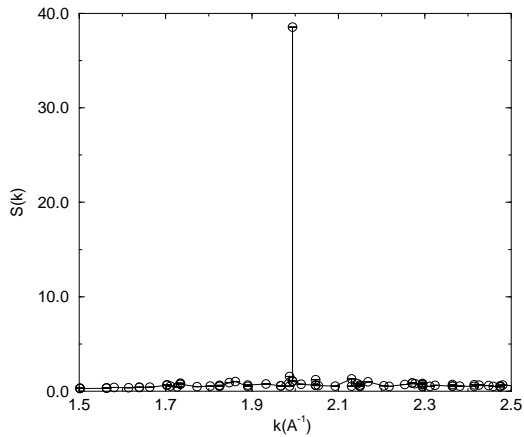}
\caption{Static structure factor $S(k)$ at the density $0.0849 \AA^{-2}$.
The simulation cell is 38.331 $\AA$ $\times$ 22.131 $\AA$ 
(72 $H_{2}$ molecules).
The sharp peak position is $k$= 1.99 $\AA^{-1}$.}
\label{fig-ic1fw}
\end{figure}

The first layer of molecular hydrogen adsorbed on graphite forms 
an incommensurate solid phase before the first layer is complete. 
We have carried out a simulation at the density
$\rho= 0.0849 \AA^{-2}$ using a simulation cell with dimensions
$x$= 9 $\sqrt{3}$ $a_{gr}$ and $y$= 9 $a_{gr}$. 
The calculated probability distribution(Fig.~\ref{fig-ic2fw}) also clearly 
shows the presence of  an  equilateral triangular solid structure 
which is not registered with the underlying graphite lattice
and it is rotated relative to the graphite lattice.
This rotation was first predicted and discussed by 
Novaco and McTague\cite{novaco-mctague}.
The angle of the incommensurate lattice relative to the graphite lattice
is approximately $5^{ \circ}$ in good agreement with the experimental
value\cite{Cui} for the above density.
The calculated static
structure factor (shown in Fig.~\ref{fig-ic1fw}) has one sharp peak 
at $k$= 1.99 $\AA^{-1}$ corresponding 
to an unregistered equilateral triangular lattice in agreement 
with the value of $k$= 1.97 $\AA^{-1}$
reported from the neutron 
diffraction experimental study\cite{Freimuth87}.

\section{Energy Calculation}

\begin{figure}[htp]
\includegraphics[width=\figwidth]{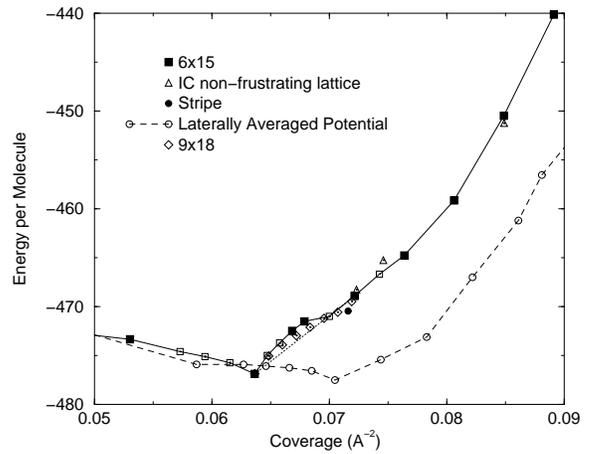}
\caption{The dashed line is the energy per hydrogen molecule on the graphite
substrate using the laterally averaged potential as a function of surface 
coverage. The solid line is the same quantity when corrugations are
included.} 
\label{fig-compare}
\end{figure}

In Fig.~\ref{fig-compare} we compare the energy per hydrogen molecule
as a function of the surface coverage for the cases where the full potential 
is used (solid line) and the case where only the
laterally averaged potential is used (dashed line). 
There are several important observations which will make based on the results
presented on the this figure. First the curve obtained using
the laterally averaged potential has a minimum at the density 
$\rho_0=0.0705 \AA^{-2}$. At this equilibrium density the system 
forms a triangular
lattice which fills the entire system and below this density the
system undergoes a phase separation and forms a solid-vapor coexistence
phase. Therefore below $\rho_0$ the energy as a function of 
density is higher for a 
finite-size system
or for a system which is forced to be uniform at this lower density. 
Above this density the system is a compressed triangular solid until
the promotion density\cite{NM}. 
The solid line corresponds to the energy per molecule when we used the 
hydrogen-molecule-graphite interaction potential with the corrugations.
The minimum in this case occurs at  $\rho_c=0.0636 \AA^{-2}$ which corresponds
to the $\sqrt{3}\times \sqrt{3}$ solid. Below that density 
the system is unstable to
formation of solid clusters and this part of the phase diagram was 
investigated in Ref.\cite{NM}. Above this density up to a density
one can notice that the presence of the substrate corrugations
moves the minimum from $\rho_0$ at much higher energy but there
is a feature at the same density produced by the $H_2$-$H_2$ interaction.
The value $\rho_0$ of the equilibrium density for a smooth substrate 
plays a significant
role even when the full potential with the substrate corrugations
is used because it is the density which minimizes the energy
due to $H_2$ molecule-molecule interaction.
Clearly due to this term which has a preference for a higher density
than the commensurate density, the solid line is not convex 
which a sign of instability of
a uniform phase. These results are obtained
on lattices which frustrate the proper periodicity of the domain walls.
If the calculations are done on lattice which can accommodate the
periodicity of these superlattice structure the energy
is lowered. As an example we also give (solid circle) the energy
obtained for the stripe state of Fig.~\ref{fig-60fw}.
Notice that the energy falls below that obtained with choice of 
lattices unfavorable for this ordered domain-wall state.  

\section{Conclusions}

Our studies of $H_2$ monolayer on graphite above 1/3 coverage clarifies
the nature of the phase diagram. The boundaries of the phase diagram
were well-known from the specific anomalies. We find that the phase 
near 1/3 coverage is the known $\sqrt{3}\times \sqrt{3}$ commensurate solid.
At higher densities the monolayer undergoes a transition
to a uniaxially compressed solid with density wave stripe-like modulations
along the same direction. This phase was first identified by LEED\cite{Cui86}
studies; while the subsequent neutron diffraction studies\cite{Freimuth87}
confirm the existence of ordered domains walls, the peak which is the 
signature of
the uniaxial compression is not seen because of the (002) reflection from
the underlying graphite lattice. In addition in both studies
the domain walls are assumed to be superheavy. As it is clear from
both our contour plots and our structure factor the walls are not 
significantly denser than the rest of the system; we find that the additional
molecules  added to the monolayer to increase the coverage above the 
commensurate coverage are used primarily to uniaxially compressed
the solid and they are only partially used to construct the walls.
Our calculated specific heat for this coverage range as a function of
 temperature has two anomalies at two temperature values.
From the calculated contour plots and the structure factor,
the lower temperature anomaly is identified as the 
transition from the ordered domain-wall solid to a solid where the
domain walls form a fluid and the higher temperature anomaly
is  caused by the total melting of the solid to a uniform fluid.

At even  higher density we find that the uniaxially compressed solid
with the ordered domain walls undergoes a transition to a triangular
solid which is incommensurate with respect to the substrate lattice and 
in addition the triangular solid of the adlayer is rotated by an 
angle of the order of $5^{\circ}$ with respect to the underlying substrate
lattice. This is in complete agreement with both the
theoretical prediction\cite{novaco-mctague} and experimental 
findings\cite{Cui86,Freimuth87}. 

\section{Acknowledgements}

This work was supported by a National Aeronautics and Space Administration
grant number NAG8-1773.

\end{document}